\documentclass[12pt,draft,titlepage]{article}
\usepackage{amssymb}
\usepackage{amsmath}
\usepackage[dvips]{graphicx}
\usepackage{epsf}
\usepackage[english]{babel}

\title{\bf Production of two gluons in the Lipatov effective
action formalism}
\author{M.A.Braun, M.Yu.Salykin and M.I.Vyazovsky\\
 Saint-Petersburg State University,\\
198504 S.Petersburg, Russia}

\def\beq{\begin{equation}}
\def\eeq{\end{equation}}
\def\noi{\noindent}

\begin{document}

\maketitle
\medskip
\noi{\Large\bf Abstract}

The one-loop diffractive amplitude
for emission of two real gluons with widely different rapidities
is studied in the Lipatov effective action formalism. It is shown that
after integration over  longitudinal momenta in the loop
the resulting expression coincides with the one
obtained by the Lipatov-Bartels formalism in transversal space
provided the same prescription is used to exclude divergent
contributions as previously proposed for emission of a single
real gluon.

\section{Introduction}
In the framework of the perturbative QCD, in the Regge kinematics,
particle interaction is described by the exchange of reggeized gluons
which emit and absorb real gluons with certain production vertexes
("Lipatov vertexes')~\cite{bfkl}. Pomeron interaction  leads to their
splitting. Emission of real gluons from split reggeized gluons is
described by vertexes introduced by J.Bartels ("Bartels vertexes')
~\cite{bartels}. Originally both type of vertexes were calculated directly
from the relevant simple Feynman diagrams in the Regge kinematics.
Later a powerful effective action formalism was proposed by L.N.Lipatov
~\cite{lipatov}, which considers reggeized and normal gluons as independent
entities from the start and thus allows to calculate all QCD diagrams
in the Regge kinematics automatically and in a systematic and self-consistent
way. However the resulting expressions are 4-dimensional and need
reduction to the final 2-dimensional transverse form.

In the paper of two co-authors of the present paper (M.A.B. amd M.I.V.)
~\cite{bravyaz} it was demonstrated that the diffractive
amplitude for the production of a real gluon calculated by means of the
Lipatov effective action, after integration over the longitudinal variables,
goes over into the expression obtained via the Lipatov and Bartels vertexes.
However in the process of reduction to the transverse form a certain
prescription had to be used to give sense to divergent integrals.

In this paper we generalize these results to a more complicated case
of production of two real gluons with a large difference in rapidity.
This case is of importance in view of the contradiction between the
results obtained by Yu.Kovchegov and K.Tuchin ~\cite{kovt}, on the one hand,
and J.Bartels, M.Salvadore and G.P.Vacca ~\cite{barsv}, on the other
for the inclusive cross-section of gluon production
in the Regge kinematics. Analysis of these results requires to compare
expressions for the two-gluon production amplitude in the Lipatov-Bartels
and dipole pictures. The study of this amplitude in the Lipatov effective
action formalism is thus a valuable test of the presently used expressions.

Our results demonstrate that the Lipatov effective action leads to the
standard expression for the two gluon production amplidude with the Lipatov
and Bartels vertexes, provided the same prescription for the longitudinal
integration is used as in ~\cite{bravyaz}.

\section{The set of diagrams}
Our purpose is to study the amplitude for the production of two gluons
in the diffractive collision on a colorless target. To simplify we shall
restrict ourselves with a case when both colliding particle are quarks.
This will introduce infrared divergence in the final intergrations over the
transferred transverse momenta, absent with the
realistic colorless participants. However our final goal is only to
obtain the amplitudes with fixed transverse variables to be able to compare
with the corresponding expression in the Lipatov-Bartels formalism.
For this particular purpose using quarks as the projectile and target
is sufficient. And it substancially reduces the number of diagrams to study.

Contributions to the process we study start at the perturbative order $g^6$,
with which we limit ourselves here.
All relevant diagrams then can be split into three groups shown in
Figs.~1,2,3. Reggeized gluons are shown by wavy lines.
The first group (\ref{Fig1}) consists of diagrams in which one
gluon (the harder) is emitted before the reggeized gluon splits into two
and the other at the splitting vertex.
The second group (\ref{Fig2}) represents diagrams in which the harder gluon
is emitted at the splitting vertex. Finally the third
group (\ref{Fig3}) is represented by digrams in which the reggeized gluons
do not split at all. Note that the contribution from the diagrams in which
the harder gluon is emitted before splitting and the softer after splitting
is equal to zero, since the splitting vertex without emission of a real
gluon vanishes due to signature conservation.

We denote the momentum of the incident quark $k$ and that of the
target quark $l$. Their final momenta are $k'$ and $l'$
respectively. We assume that $k_-=k_\perp=l_+=l_\perp=0$.
The momenta of the emitted gluons are $p_1$ and $p_2$
with $p_{1+}>>p_{2+}$. For all longitudinal components we use
the definition $a_{\pm}=a_{0}\pm a_{3}$, so that
$ab=\frac{1}{2}a_{+}b_{-}+\frac{1}{2}a_{-}b_{+}+(ab)_{\perp}$.

In order to have the uniform notations for all diagrams we used
the following definitions of various transferred momenta:
\beq
q=k-k',\ \
q_{2}=q-p_{1}-q_{1},\ \
q_{3}=q-p_{1}-p_{2}-q_{1},\ \
q_{4}=q-p_{1},\ \
q_{5}=q-q_{1},\ \
q_{6}=q-q_{2}\, ,
\label{1-6}
\end{equation}
where $q_{1}$ is a loop momentum.
In the Regge kinematics we have:
\[
\sqrt{s}=k_{+}\approx k'_{+}\gg p_{1+}\sim q_{+}\gg p_{2+}
\sim q_{4+}\gg l'_{+}\]
\[
\sqrt{s}=l_{-}\approx l'_{-}\gg p_{2-}\gg p_{1-}
\sim -q_{4-}\gg k'_{-}
\]
\beq
\label{9}
q_{5+}\ll \sqrt{s}, q_{6+}\ll\sqrt{s}, q_{1-}\ll\sqrt{s}, q_{2-}\ll\sqrt{s}
\, .
\eeq
We recall that in the Regge kinematics non-zero tranversal momentum
components are assumed to be much smaller than longitudinal ones.

\begin{figure}[h]
\leavevmode \centering{\epsfysize=0.3\textheight\epsfbox{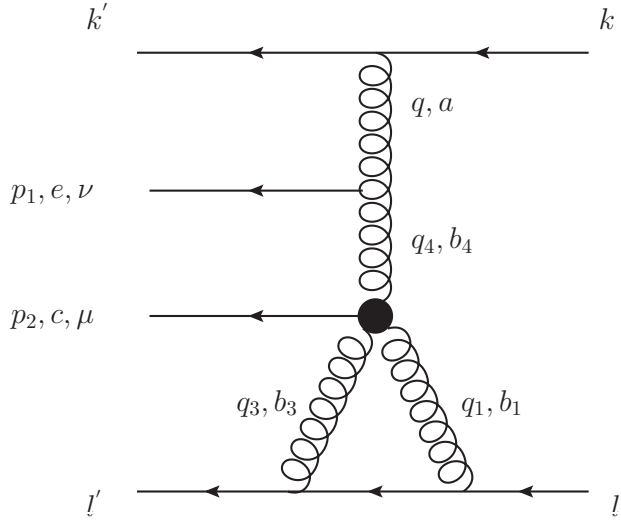}}
\caption{Diagram of the type 1}
\label{Fig1}
\end{figure}

\section{Diagram of Fig. \ref{Fig1}}
The only diagram of type 1 is shown in Fig. \ref{Fig1}.
We denote the wave functions of the projectile and target as
$\bar{u}(k')$, $u(k)$ and $\bar{w}(l')$, $w(l)$
correspondingly. So the factors describing the projectile and target
quarks are correspondingly
\begin{equation}
\label{11}
ig\bar{u}(k')\frac{\gamma_{+}}{2}t^{a}u(k)
\end{equation}
and
\begin{equation}
\label{12}
(ig)^2\bar{w}(l')t^{b_{3}}\frac{\gamma_{-}}{2}
\frac{i(\hat{l}+\hat{q_{1}})}{(l+q_{1})^2+i0}
\frac{\gamma_{-}}{2}t^{b_{1}}w(l)\, .
\end{equation}
We can use (\ref{9}) to simplify
\begin{equation}
\label{13}
\gamma_{-}(\hat{l}+\hat{q_{1}})\gamma_{-}=
\frac{1}{2}\gamma_{-}(l_{-}+q_{1-})\gamma_{+}\gamma_{-}=
2\gamma_{-}(\sqrt{s}+q_{1-})\approx
2\sqrt{s}\gamma_{-}\, .
\end{equation}
We shall be interested in the diffraction process when the target
does not change its colour and so the  $t$-channel coupled to
the target is colourless. So we introduce a projector onto
the colourless target state
\beq
P_{b_1b_3|b'_1b'_3}=\frac{\delta_{b_1b_3}\delta_{b'_1b'_3}}{N_c^2-1}\, .
\label{19}
\eeq
Acting on the target quark colours it gives a factor
\beq
\frac{1}{N_c^2-1}\delta_{b'_1b'_3}t^{b'_3}t^{b'_1}=\frac{1}{2N_c}\, .
\label{colour}
\eeq

The diagram of Fig. \ref{Fig1} is formed by the vertexes already
studied previously. The lower vertex is
the Reggeon $\to$ 2 Reggeons + Particle ("effective") vertex
obtained in ~\cite{bravyaz}:

\begin{equation}
\begin{split}
\frac{ig^{2}f^{b_{3}cd}f^{b_{4}b_{1}d}}{(q_{4}-q_{1})^{2}}
\left[
q_{4+}(q_{4}\varepsilon^{*}_{2})_{\bot}+
\frac{q_{4}^{2}}{q_{1-}}
\left(-((q_{4}-q_{1})\varepsilon^{*}_{2})_{\bot}
+\frac{(q_{4}-q_{1})^{2}}{p_{2\bot}^{2}}(p_{2}\varepsilon^{*}_{2})_{\bot}
\right)\right]
\\
+\frac{ig^{2}f^{b_{1}cd}f^{b_{4}b_{3}d}}{(q_{4}-q_{3})^{2}}
\left[
q_{4+}(q_{4}\varepsilon^{*}_{2})_{\bot}+\frac{q_{4}^{2}}{q_{3-}}
\left(-((q_{4}-q_{3})\varepsilon^{*}_{2})_{\bot}
+\frac{(q_{4}-q_{3})^{2}}{p_{2\bot}^{2}}(p_{2}\varepsilon^{*}_{2})_{\bot}
\right)\right]\, .
\label{15}
\end{split}
\end{equation}
The third term in each square brackets is the contribution
of the so-called "induced" vertex  which is given by expansion of
the $P$-exponential in the effective action~\cite{lipatov}.

The upper vertex is the well-known
Reggeon $\to$ Reggeon $+$ particle ("Lipatov") vertex, which we write as:
\begin{equation}
\label{16}
igf^{ab_{4}e}q_{\bot}^{2}\Big(L_{\perp}(q,p_1)\varepsilon^{*}_{1}\Big)\, ,
\end{equation}
where we define the transverse vector
\beq
L_\nu(q,p)=\frac{q_{\bot\nu}}{q^2_{\bot}}-\frac{p_{\bot\nu}}{p^2_{\bot}}\, .
\label{16a}
\eeq

The effective vertex consists of two parts proportional to
$f^{b_3cd}f^{b_4b_1d}$ and
and $f^{b_{1}cd}f^{b_{4}b_{3}d}$,
each containing three terms.
In both cases the convolution of the vertex  colour factors with the
target colour factor (\ref{colour}) gives the final overall colour factor
\begin{equation}
\label{20}
\frac{1}{2N_c}f^{ab_{4}e}f^{b_{1}cd}f^{b_{4}b_{1}d}t^a=
\frac{1}{2}f^{aec}t^{a}\, .
\end{equation}

To reduce the contribution of the diagram to the 2-dimensional form we have
to integrate over the longitudinal variables in the loop.
This integration
does not involve the four reggeon propagators as
they are purely transversal
\begin{equation}
\label{13a}
D^{ab}(q)=-i\frac{2\delta_{ab}}{q^{2}_{\bot}}
\end{equation}
and thus contribute a totally transverse factor
\begin{equation}
\label{14}
\frac{16}{q^2_{\bot}q^2_{4\bot}q^2_{1\bot}q^2_{3\bot}}\, .
\end{equation}

The effective vertex generates three kinds of terms including
longitudinal components proportional to
\begin{equation}
\label{21}
q_{4+}(q_{4}\varepsilon^{*}_{2})_{\bot}\, ,
\end{equation}

\begin{equation}
\label{22}
\frac{q_{4}^{2}}{q_{i-}}(-((q_{4}-q_{i})
\varepsilon^{*}_{2})_{\bot}),\ \  i=1,3
\end{equation}
and
\begin{equation}
\label{24}
\frac{q_{4}^{2}}{q_{i-}}\frac{(q_{4}-q_{i})^{2}}
{p_{2\bot}^{2}}(p_{2}\varepsilon^{*}_{2})_{\bot},\ \ i=1,3\, .
\end{equation}

Combined with the denominator from the quark propagator
the first two terms lead to the longitudinal
integrals of two  forms
\begin{equation}
J_1(k_1,k_2)=
\frac{1}{2i}
\int\frac{dq_{1-}}{2\pi}\int\frac{dq_{1+}}{2\pi}
\frac{1}{(k_1^2+i0)(k_2^2+i0)}
\label{form1}
\end{equation}
and
\begin{equation}
J_2(k,k_1,k_2)=
\frac{1}{2i}
\int\frac{dq_{1-}}{2\pi}\int\frac{dq_{1+}}{2\pi}\frac{1}{k_-}
\frac{1}{(k_1^2+i0)(k_2^2+i0)}\ .
\label{form2}
\end{equation}
where $k$, $k_1$ and $k_2$ are some linear functions of the
integration momentum $q_1$.

The first terms, proportional to (\ref{21}),
lead to the integrals
\beq
\label{26}
I_1=J_1(q_4-q_1,l+q_1)
\eeq
and
\begin{equation}
\label{27}
I_2=J_1(q_4-q_3,l+q_1)\, .
\end{equation}
The second terms (\ref{22}) combined with the target quark denominator
lead to the integrals
\begin{equation}
\label{28}
I_3=J_2(q_1,q_4-q_1,l+q_1)
\end{equation}
and
\begin{equation}
\label{29}
I_4=J_2(q_3, q_4-q_3,l+q_1)\, .
\end{equation}

The third terms (\ref{24}) combined with the target quark propagator
give the integrals
\begin{equation}
\frac{1}{2i}
\int\frac{dq_{1-}}{2\pi}
\int\frac{dq_{1+}}{2\pi}
\frac{1}{q_{1-}}\frac{1}{(l+q_1)^2 +i0}
\label{26a}
\end{equation}
and
\begin{equation}
\frac{1}{2i}
\int\frac{dq_{1-}}{2\pi}\int\frac{dq_{1+}}{2\pi}
\frac{1}{q_{3-}}\frac{1}{(l+q_1)^2 +i0}\ .
\label{26b}
\end{equation}
In these formulas $q_3=q-p_1-p_2-q_1$.

The first four integrals (\ref{26}-\ref{29}) are calculated in the Appendix.
The last integrals (\ref{26a}), (\ref{26b}) are formally divergent.
The same integrals were also found  in the simpler case of the
single gluon production in ~\cite{bravyaz}. There it was noted that
if a prescription is imposed to calculate the integral in
the principal value sense then the integral vanishes and the result turns
out to be in agreement with the standard Lipatov-Bartels approach in terms
of ordinary Feynman diagrams. Relying on this conclusion we also in this
study impose the same rule of calculation and consequently
neglect these integral altogether.

The results  found in the Appendix for the sum of the first two integrals
is
\begin{equation}
\label{30}
I_1+I_2=
\frac{i}{4 q_{4+}\sqrt{s}}\, ,
\end{equation}
attaching the rest factor from the effective vertex
we find the total contribution to the diagram from the terms (\ref{21}) as
\begin{equation}
\label{33}
\frac{i(q_{4}\varepsilon^{*}_{2})_{\bot}}{4\sqrt{s}}\ .
\end{equation}

For the second terms, the factors (\ref{22}) are different for
the two parts of the effective vertex. If we change the variable
of the loop integration $q_{1}\to q_{3}=q-p_{1}-p_{2}-q_{1}$
only in the contribution of the second part, then the factors (\ref{22})
become equal and we need to calculate the sum of integrals $I_3$
and $I_4=J_2(q_1, q_4-q_1,l'-q_1)$. The sum of integrals is found to be
\begin{equation}
I_3+I_4=\frac{i}{4\sqrt{s}(q_{4}-q_{1})^2_{\bot}}\ .
\label{33a}
\end{equation}
Combined with the rest factor from the effective vertex
they give the contribution from the terms (\ref{22}) as
\begin{equation}
\label{34}
-\frac{iq_{4\bot}^{2}}{4\sqrt{s}(q_{4}-q_{1})^2_{\bot}}
((q_{4}-q_{1})\varepsilon^{*}_{2})_{\bot}\ .
\end{equation}

Summing (\ref{33}) and (\ref{34}) we find the final result
for the diagram of type 1 in the form
\begin{equation}
\label{35}
g^{6}\cdot\bar{u}(k')\gamma_{+}\Delta_{1}u(k)\cdot
\bar{w}(l')\gamma_{-}w(l)\, ,
\end{equation}
where
\begin{equation}
\label{36}
\Delta_{1}=\frac{i}{2}f^{aec}t^{a}\cdot
\int \frac{d^{2}q_{1\bot}}{(2\pi)^2}\frac{1}{q^2_{1\bot}}
\frac{1}{q^2_{3\bot}}\left(L_{\bot}(q,p_{1})\varepsilon^{*}_{1}\right)
\left(B_{\bot}(q_{4},q_2)\varepsilon^{*}_{2}\right)\, .
\end{equation}
Here we denoted
\begin{equation}
\label{37}
B_{\nu}(q,p)=\frac{q_{\bot\nu}}{q^{2}_{\bot}}-
\frac{p_{\bot\nu}}{p^{2}_{\bot}}
\end{equation}
with
\begin{equation}
q_{2}=q_{4}-q_{1}=q-p_{1}-q_{1},\ \  q_{3}=q-p_{1}-p_{2}-q_{1}\, .
\end{equation}
This is the momentum part of the
well-known Bartels vertex \cite{bartels} expressed
in terms of this part for the Lipatov vertex.
Expression (\ref{36}) is exactly the one which is found for the configuration
of the diagram in Fig. \ref{Fig1} in the Lipatov-Bartels
formalism using the transverse space approach from the start.

\begin{figure}[t]
\leavevmode \centering{\epsfysize=0.25\textheight\epsfbox{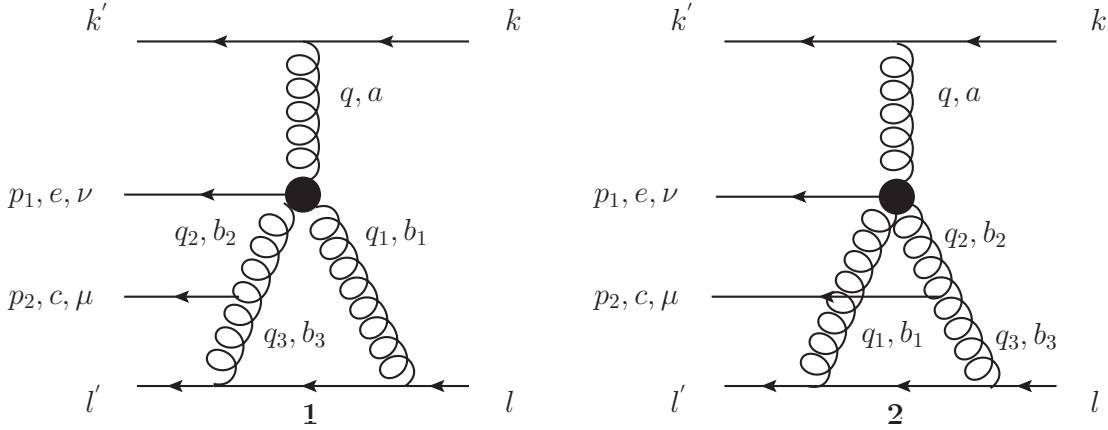}}
\caption{Diagrams of the type 2}
\label{Fig2}
\end{figure}

\section{Diagrams of Fig. \ref{Fig2}}
The two diagrams of type 2 are shown in Fig. \ref{Fig2}.
The softer gluon is now
emitted from inside the loop.
The structure of the diagrams   is similar to the previous case
except that the effective vertex has the larger rapidity than the
Lipatov vertex.

The target factor for the diagram Fig. \ref{Fig2}.1 is
\begin{equation}
\label{39}
\bar{w}(l')t^{b_{3}}\frac{ig\gamma_{-}}
{2}\frac{i(\hat{l}+\hat{q_{1}})}{(l+q_{1})^2+i0}
\frac{ig\gamma_{-}}{2}t^{b_{1}}w(l)
\end{equation}
and for the diagram Fig. \ref{Fig2}.2 is
\begin{equation}
\label{39a}
\bar{w}(l')t^{b_{1}}\frac{ig\gamma_{-}}
{2}\frac{i(\hat{l'}-\hat{q_{1}})}{(l'-q_{1})^2+i0}
\frac{ig\gamma_{-}}{2}t^{b_{3}}w(l)\, .
\end{equation}
Similar to (\ref{13}):
\begin{equation}
\label{40}
\gamma_{-}(\hat{l'}-\hat{q_{1}})\gamma_{-}
\approx
2\sqrt{s}\gamma_{-}\ .
\end{equation}
Reggeon propagators give a factor
\begin{equation}
\label{41}
\frac{16}{q^2_{\bot}q^2_{1\bot}q^2_{2\bot}q^2_{3\bot}}\ .
\end{equation}

For the first diagram Fig. \ref{Fig2}.1
the effective vertex is
\begin{eqnarray}
\frac{ig^{2}f^{b_{1}ed}f^{ab_{2}d}}{(q-q_{2})^{2}}
\left[
q_{+}(q\varepsilon^{*}_{1})_{\bot}+\frac{q^{2}}{q_{2-}}
\left(-((q-q_{2})\varepsilon^{*}_{1})_{\bot}
+\frac{(q-q_{2})^{2}}{p_{1\bot}^{2}}(p_{1}\varepsilon^{*}_{1})_{\bot}\right)
\right]
\nonumber
\\
+
\frac{ig^{2}f^{b_{2}ed}f^{ab_{1}d}}{(q-q_{1})^{2}}
\left[
q_{+}(q\varepsilon^{*}_{1})_{\bot}+\frac{q^{2}}{q_{1-}}
\left(-((q-q_{1})\varepsilon^{*}_{1})_{\bot}
+\frac{(q-q_{1})^{2}}{p_{1\bot}^{2}}(p_{1}\varepsilon^{*}_{1})_{\bot}\right)
\right]\ .
\label{42}
\end{eqnarray}
and the Lipatov vertex is
\begin{equation}
\label{43}
igf^{b_{2}b_{3}c}q_{2\bot}^{2}
\left(L_{\bot}(q_2,p_2)\varepsilon^{*}_{2}\right)=
igf^{b_{2}b_{3}c}q_{2\bot}^{2}
\left(
\frac{(q_{2}\varepsilon^{*}_{2})_{\bot}}{q^{2}_{2\bot}}
-\frac{(p_{2}\varepsilon^{*}_{2})_{\bot}}{p^{2}_{2\bot}}
\right)\ .
\end{equation}

For the second diagram Fig. \ref{Fig2}.2  the effective vertex
is the same (\ref{42}), since it is invariant under interchange
of the two lower reggeons,
and the  Lipatov's vertex is also (\ref{43}).

The projection of the reggeons coupled to the target
onto the colorless state supplies the same factor as for the
diagram of Fig. \ref{Fig1} (\ref{19}). Its convolution
with other color factors for both digarams of Fig. \ref{Fig2}
however gives different results for the two parts of the effective vertex.
For the first part we get
\begin{equation}
\label{47}
\frac{1}{2N_{c}}f^{b_{1}ed}f^{ab_{2}d}f^{b_{2}b_{1}c}t^a
=-\frac{1}{4}f^{aec}t^{a}
\end{equation}
and for the second part we get the opposite sign
\begin{equation}
\label{48}
\frac{1}{2N_{c}}f^{b_{2}ed}f^{ab_{1}d}f^{b_{2}b_{1}c}t^a
=\frac{1}{4}f^{aec}t^{a}\, .
\end{equation}

Further calculations are quite similar to those for
the diagram on Fig. \ref{Fig1}, except that now we have to consider
the two parts of the effective vertex separately. In the following we
choose the integration variable to be $q_{1}$.

Consider the diagram Fig. \ref{Fig2}.1. The first terms in the two parts
of the effective vertex lead to the integrals respectively
\beq
I_5=J_1(q-q_2,l+q_1)=0
\eeq
and
\beq
I_6=J_1(q-q_1,l+q_1)=\frac{i}{4 q_{+}\sqrt{s}}\, .
\eeq
Notice that $I_5$ enters with the color factor (\ref{47})
and $I_6$ does with the color factor (\ref{48}).
For the second diagram Fig. \ref{Fig2}.2 the integrals for the
first terms of the effective vertex are
\beq
I_7=J_1(q-q_2,l'-q_1)=\frac{i}{4 q_{+}\sqrt{s}}
\eeq
and
\beq
I_8=J_1(q-q_1,l'-q_1)=0
\eeq
and the color factors are (\ref{47}) and (\ref{48})
respectively.
Since the contributions of the two non-zero integrals
enter with the opposite sign, in the sum of the two diagrams
in Fig. \ref{Fig2} we get zero.

The last terms in the effective vertex again formally diverge. Using our
prescription of the principal value integration we put them to zero.
So in the end only the second terms in the effective vertex give non-zero
contribution.

For the first diagram in Fig. \ref{Fig2} they lead to the longitudinal
integrals
$
I_9=J_2(q_2,q-q_2,l+q_1)
$
and
$
I_{10}=J_2(q_1,q-q_1,l+q_1).
$
For the second diagram in Fig. \ref{Fig2} one has to change
$l+q_1\to l'-q_1$:
$
I_{11}=J_2(q_2,q-q_2,l'-q_1)
$
and
$
I_{12}=J_2(q_1,q-q_1,l'-q_1).
$

The integrals are similar to that
we have calculated for diagram in Fig. \ref{Fig1}.
Color factor (\ref{47}) corresponds to $I_9$ and $I_{11}$.
Summing them we obtain
\begin{equation}
-\frac{1}{4}f^{aec}t^{a}(I_9+I_{11})=
-\frac{1}{4}f^{aec}t^{a}\frac{i}{4 (q-q_{2})^{2}_{\bot}}\, .
\label{48d}
\end{equation}
Similarly, the second part gives
\begin{equation}
\frac{1}{4}f^{aec}t^{a}(I_{10}+I_{12})=
\frac{1}{4}f^{aec}t^{a}\frac{i}{4 (q-q_{1})^{2}_{\bot}}\, .
\label{48e}
\end{equation}

Taking in account that vertex (\ref{43}) is the same for both diagrams
we obtain for the sum of diagrams Fig.~\ref{Fig2}:
\begin{equation}
\label{56}
g^{6}\cdot\bar{u}(k')\gamma_{+}\Delta_{2}u(k)
\cdot\bar{w}(l')\gamma_{-}w(l)\, ,
\end{equation}
where
\begin{eqnarray}
\label{57}
\Delta_{2}=\frac{i}{4}f^{aec}t^{a}\cdot\int
\frac{d^{2}q_{1\bot}}{(2\pi)^2}\frac{1}{q^2_{1\bot}}
\frac{1}{q^2_{3\bot}}
\left[
(B_{\bot}(q,q_{5})\varepsilon^{*}_{1})
-(B_{\bot}(q,q_{6})\varepsilon^{*}_{1})
\right]
\left(L_{\bot}(q_{2},p_{2})\varepsilon^{*}_{2}\right)\, .
\end{eqnarray}
The definition of $B_{\nu}$ was made in (\ref{37}).
This result is also corresponding to the Lipatov-Bartels formalism.

\begin{figure}[t]
\leavevmode \centering{\epsfysize=0.8\textheight\epsfbox{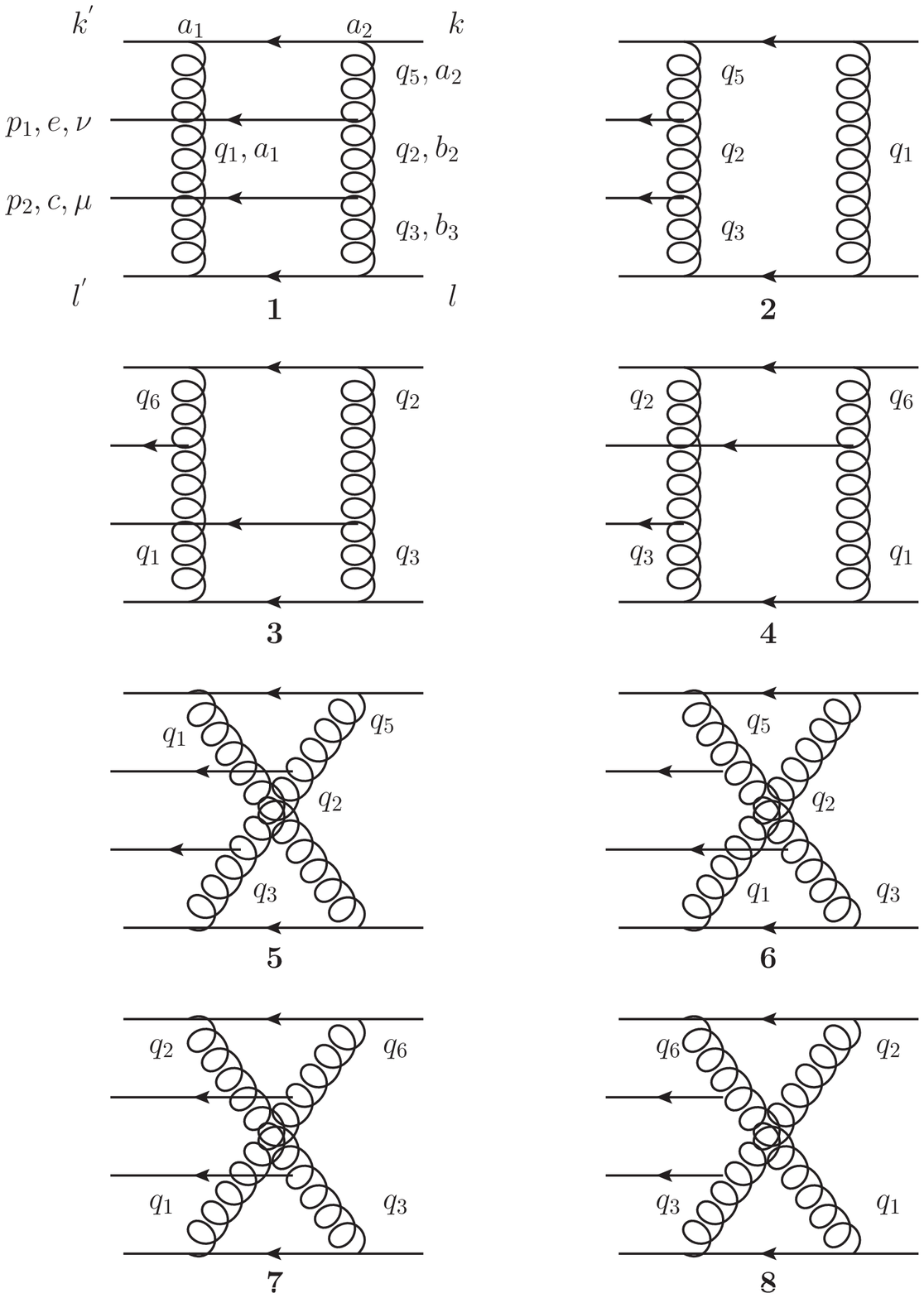}}
\caption{Diagrams of the type 3}
\label{Fig3}
\end{figure}

\section{Diagrams of Fig. \ref{Fig3}}
The diagrams on Fig (\ref{Fig3}) divide into two parts:
with emission of the two gluons  from the same reggeon
(diagrams 1,2,5 and 6) and from different reggeons
(diagrams 3,4,7 and 8). These two parts have different structures of
the Lipatov vertices.
Consider diagram 1.
Factors coupled with the target and projectile quarks are:
\begin{equation}
\label{58}
\bar{w}(l')t^{a_{1}}\frac{ig\gamma_{-}}
{2}\frac{i(\hat{l'}-\hat{q_{1}})}{(l'-q_{1})^2+i0}
\frac{ig\gamma_{-}}{2}t^{b_{3}}w(l)
\end{equation}
and
\begin{equation}
\label{59}
\bar{u}(k')t^{a_{1}}\frac{ig\gamma_{+}}
{2}\frac{i(\hat{k'}+\hat{q_{1}})}{(k'+q_{1})^2+i0}
\frac{ig\gamma_{+}}{2}t^{a_{2}}u(k)\, .
\end{equation}
As before we simplify
\begin{equation}
\label{60}
\gamma_{-}(\hat{l'}-\hat{q_{1}})\gamma_{-}\approx 2\sqrt{s}\gamma_{-}\ ,
\end{equation}
\begin{equation}
\label{61}
\gamma_{+}(\hat{k'}+\hat{q_{1}})\gamma_{+}\approx 2\sqrt{s}\gamma_{+}\ .
\end{equation}
The reggeon propagators are:
\begin{equation}
\label{62}
\frac{16}{q^2_{1\bot} q^2_{2\bot}
q^2_{3\bot} q^2_{5\bot}}\, .
\end{equation}
The two Lipatov vertices are:
\begin{equation}
\label{63}
igf^{a_{2}b_{2}e}q_{5\bot}^{2}
\left(
\frac{(q_{5}\varepsilon^{*}_{1})_{\bot}}{q^{2}_{5\bot}}
-\frac{(p_{1}\varepsilon^{*}_{1})_{\bot}}{p^{2}_{1\bot}}
\right)\, ,
\end{equation}
\begin{equation}
\label{64}
igf^{b_{2}b_{3}c}q_{2\bot}^{2}
\left(
\frac{(q_{2}\varepsilon^{*}_{2})_{\bot}}{q^{2}_{2\bot}}
-\frac{(p_{2}\varepsilon^{*}_{2})_{\bot}}{p^{2}_{2\bot}}
\right)\, .
\end{equation}
After the projection of the reggeons coupled to the target
onto the colorless state we obtain the following color structure:
\begin{equation}
\label{65}
\frac{1}{2N_{c}}f^{a_{2}b_{2}e}f^{b_{2}a_{1}c}t^{a_{1}}t^{a_{2}}\ .
\end{equation}
The diagram 5 differs from this one only in the target quark propagator
in which $l'-q_1\to l+q_1$.
The diagrams 2 and 6 have a different color structure
\begin{equation}
\label{66}
\frac{1}{2N_{c}}f^{a_{2}b_{2}e}f^{b_{2}a_{1}c}t^{a_{2}}t^{a_{1}}\ .
\end{equation}

It is convenient  to combine momentum parts of these four diagrams.
In order to do it we split
$t^{a_{1}}t^{a_{2}}$
into symmetric and antisymmetric parts:
\begin{equation}
\label{68}
t^{a_{1}}t^{a_{2}}=\frac{1}{2}
 \{t^{a_1},t^{a_2}\}+\frac{1}{2}[t^{a_{1}},t^{a_{2}}]
\end{equation}
to obtain colour factors
\begin{equation}
\label{69}
\frac{1}{4N_{c}}f^{a_{2}b_{2}e}f^{b_{2}a_{1}c}[t^{a_{1}},t^{a_{2}}]
=\frac{i}{8}f^{aec}t^{a}
\end{equation}
\begin{equation}
\label{70}
\frac{1}{4N_{c}}f^{a_{2}b_{2}e}f^{b_{2}a_{1}c}\{t^{a_{1}},t^{a_{2}}\}
=-\frac{1}{4N_c}\delta^{ce}-\frac{1}{8}d^{aec}t^{a}\ .
\end{equation}

Thus for the antisymmetric part we use (\ref{69}) with an extra minus
sign for diagrams 2 and 6. For the total symmetric part we
use (\ref{70}) for all four diagrams.

The longitudinal integrals for diagrams 1, 2, 5 and 6 are
correspondingly
\begin{equation}
I_{13}=J_1(k'+q_{1},l'-q_1)=\frac{i}{4s}\, ,
\end{equation}
\begin{equation}
I_{14}=J_1(k-q_{1},l+q_1)=\frac{i}{4s}\, ,
\end{equation}
\begin{equation}
I_{15}=J_1(k'+q_{1},l+q_1)=0\, ,
\end{equation}
\begin{equation}
I_{16}=J_1(k-q_{1},l'-q_1)=0\, .
\end{equation}

As we observe, the antisymmetric part completely vanishes.
In the symmetric part for the sum of diagrams 1, 2, 5 and 6
we obtain
\beq
\label{75}
2i\left(-\frac{\delta^{ce}}{4N_{c}}-\frac{d^{aec}}{8}t^{a}\right)\cdot\int
\frac{d^{2}q_{1\bot}}{(2\pi)^2}\frac{1}{q^2_{1\bot}}
\frac{1}{q^2_{3\bot}}
\left(L_{\bot}(q_{5},p_{1})\varepsilon^{*}_{1}\right)
\left(L_{\bot}(q_{2},p_{2})\varepsilon^{*}_{2}\right)\, .
\eeq

Calculation of diagrams 3, 4, 7 and 8 is completely similar.
The total result for all diagrams in Fig. \ref{Fig3} is
\begin{equation}
\label{76}
g^{6}\cdot\bar{u}(k')\gamma_{+}\Delta_{3}u(k)\cdot
\bar{w}(l')\gamma_{-}w(l)\, ,
\end{equation}
where
\begin{eqnarray}
\label{77}
\Delta_{3}=2i\left(-\frac{1}{4N_{c}}\delta^{ce}-\frac{1}{8}d^{aec}t^{a}\right)
\cdot
\int\frac{d^{2}q_{1\bot}}{(2\pi)^2}\frac{1}{q^2_{1\bot}}
\frac{1}{q^2_{3\bot}}
\left(L_{\bot}(q_{2},p_{2})\varepsilon^{*}_{2}\right)
\nonumber
\\
\left[
(L_{\bot}(q_{5},p_{1})\varepsilon^{*}_{1})
-(L_{\bot}(q_{6},p_{1})\varepsilon^{*}_{1})
\right]\, .
\end{eqnarray}
This  expression is exactly the one obtained in the standard
Lipatov-Bartels transverse space approach .


\section{Conclusions}

Using the Lipatov effective field theory we have generalized the
results of ~\cite{bravyaz} to the case when
in the diffractive process two real soft gluons are emiited
with large distance between their rapidities
The Reggeon $\to$ 2 Reggeons+Particle vertex
involved in the process was taken from ~\cite{bravyaz}.
The found general structure of the amplitude corresponds to what
has been known from the direct calculation of
standard Feynman diagrams. To check the full correspondence we  performed
longitudinal integrations. The encountered difficulties are the same
as with the single gluon emission. They require imposition of a
certain rule, which reduces to taking certain integrals in
the prinicpal value recipe.  With this rule obeyed, the found
expression for the production amplitude completely coincides with the
one obtained by using Lipatov and Bartels vertexes in the
transversal space from the start.

It however remains to be seen if this result is true when the target
changes its colour. Such a  process is an important part of the
inclusive soft gluon production, which is now under careful study in
view of the contradiction betwen the results found in the Lipatov-Bartels
and dipole pictures, mentioned in the introduction.
We leave this problem for future studies.

\section{Acknowledgements}
This work has been partially supported by grants RNP 2.1.1/1575
of Education and Science Ministry of Russia and RFFI 09-02-01327a.

\section{Appendix. Calculation of longitudinal integrals}
The typical longitudinal integral of the form (\ref{form1}) is
\begin{equation}
\label{79}
I_1=J_1(q_4-q_1,l+q_1)=
\frac{1}{2i}\int\frac{dq_{1-}}{2\pi}\int\frac{dq_{1+}}{2\pi}
\frac{1}{[(q_{4}-q_{1})^2+i0]}\frac{1}{[(l+q_{1})^{2}+i0]}\ .
\end{equation}
The standard procedure to calculate similar integrals is to use
that the longitudinal components of the reggeon momentum $q_{1\pm}$
can be neglected as compared to large longitudinal components
of the particles to which the reggeon is coupled, that is $q_{1+}$
is to be neglected as compared to $q_{4+}$ and $q_{1-}$ is to be neglected
as compared to $l_-$. Should we follow this procedure,  integral $I_1$
will factorize into two independent integrals over $q_{1+}$ and $q_{1-}$,
but both of them will be divergent at large $q_{1\pm}$. Below we shall
demonstrate that this procedure still can be applied not to separate integrals
like (\ref{79}) but to the sum of integrals coming from the direct and crossed
terms in our expression and also somewhat transformed to achieve convergence.
To be able to calculate separate integrals of our type we recur to a slightly
different procedure, in which the condition that $q_{1\pm}$ are small is imposed
not from the start but after integration in one of the longitudinal momenta.
Of course our procedure is fully equivalent to the standard one applied to
convergent integrals and gives identical results.

As a function of $q_{1+}$ the integrand has two poles
\begin{equation}
q_{1+}=q_{4+}-\frac{(q_{4}-q_{1})^{2}_{\bot}+i0}{q_{1-}}
\label{pole1}
\end{equation}
and
\begin{equation}
q_{1+}=-\frac{(q_{1}+l)^{2}_{\bot}+i0}{q_{1-}+l_-}\ .
\label{pole2}
\end{equation}
A non-zero result is obtained only if
the two poles in $q_{1+}$ are on the opposite sides from the real axis.
It determines the limits of the integration over $q_{1-}$. In the Regge
kinematics in any case $q_{1-}<<l_-$ so that the limits are
\begin{equation}
\label{82}
-l_{-}<q_{1-}<0\, .
\end{equation}
Thus taking the residue at (\ref{pole1}) we get an integral over $q_{1-}$
\beq
\label{83}
\frac{1}{4\pi}\int_{-l_{-}}^{0}dq_{1-}\frac{1}{D_1}\ ,
\eeq
where
\begin{equation}
D_1=q_{1-}^{2}q_{4+}+q_{1-}(l_{-}q_{4+}-
(q_{4}-q_{1})^{2}_{\bot}+q_{1\perp}^{2})-l_{-}(q_{4}-q_{1})^{2}_{\perp}-i0\, .
\label{denom}
\end{equation}

The integral over $q_{1-}$ can  be directly calculated as it stands.
However such
calculation is incorrect, since it does not take into account the
kinematical conditions which are to be fulfilled for the propagating
reggeons. In fact we have to require that both longitudinal components
of the reggeon momenta are small as compared with the transversal components.
Otherwise the longitudinal momenta have to be kept
in the reggeon propagator and, if large, will correspond to
the kinematics quite different to the Regge one.
So  we have to restrict integration in (\ref{79})
to the region
\beq     |q_{1+}q_{1-}|<<|q_{1\perp}^2|\, .
\label{cond}
\eeq
In our case from (\ref{pole1}) we have
\beq
q_{1+}q_{1-}=q_{4+}q_{1-}-(q_4-q_1)_{\perp}^2\, ,
\eeq
so that condition (\ref{cond}) transforms into
\beq
|q_{4+}q_{1-}-(q_4-q_1)_{\perp}^2|<<|q_{1\perp}^2| \ .
\label{cond1}
\eeq
This implies that the integration in $q_{1-}$
is to be restricted to a narrow interval  around the point
where the left-hand side of (\ref{cond1}) vanishes.

With small values of $q_{1-}$ of this order,
we have to drop here all terms except by those which contain a large
factor $l_-$, so that we get
\beq
D_1(q_{1-})=l_-\Big(q_{4+}q_{1-}-(q_4-q_1)_{\perp}^2 -i0 \Big)\ ,
\eeq
and according to (\ref{cond1}) the integration should go in a small
interval around the point where $D_1(q_{1-})=0$. This means that in fact
\beq
\frac{1}{D_1(q_{1-})}=
\frac{i\pi}{l_-}\delta\Big(q_{4+}q_{1-}-(q_4-q_1)_{\perp}^2\Big)\ .
\eeq
Notice that $q_{4+}\approx p_{2+}>0$.
Since $(q_4-q_1)_{\perp}^2<0$ and in the integration region also
$q_{4+}q_{1-}<0$ the $\delta$-function gives a nonzero contribution and
we obtain
\beq
I_1=\frac{i}{4 l_- q_{4_+}}\ .
\label{i1}
\eeq

Now take  integral (\ref{27}). First changing the integration variable
to $q_3$ and then redenoting it as $q_1$ we find
\begin{equation}
\label{80}
I_2=J_1(q_4-q_1,l'-q_1)=
\frac{1}{2i}\int\frac{dq_{1-}}{2\pi}\int\frac{dq_{1+}}{2\pi}
\frac{1}{[(q_{4}-q_{1})^2+i0]}\frac{1}{[(l'-q_{1})^{2}+i0]}\ .
\end{equation}
The two poles in $q_{1+}$ are now the old one (\ref{pole1}) and
\beq
q_{1+}=l'_{+}+\frac{(l'-q_1)_{\perp}^2+i0}{l'_- -q_{1-}} \ .
\eeq
Now the integration region in $q_{1-}$ is
\beq
0<q_{1-}<l'_- \ .
\eeq
We get an integral
\beq
\label{83a}
I_2=-\frac{1}{4\pi}\int_{0}^{l'_-}dq_{1-}\frac{1}{D_2}\, ,
\eeq
where
\beq
D_2(q_{1-})=q_{1-}^{2}(q_{4+}-l'_+)-q_{1-}(l'_{-}q_{4+}+
(q_{4}-q_{1})^{2}_{\bot}-(l'-q_{1})^{2}_{\bot}
+{l'}_{\perp}^2)+l'_{-}(q_{4}-q_{1})^{2}_{\bot}+i0
\eeq
and the minus sign is due to the fact that the pole (\ref{pole1})
now lies in the upper half plane.
According to our estimates, in the assumed kinematical
conditions only the term in $D_2$ which
contains $q_{1-}$ multiplied by $l'_- q_{4+}$ is to be kept,
so that
\beq
D_2(q_{1-})=l'_-\Big((q_4-q_1)^2-q_{1-}q_{4_+} +i0\Big)
\label{d2}
\eeq
and according to (\ref{cond1}) we have to integrate over $q_{1-}$
in the small interval around the point where $D_2(q_{1-})=0$.
But now in (\ref{d2}) the right-hand side never vanishes, since
in the brackets both terms are negative in the integration region.
So we find
\beq
I_2=0
\label{i2}
\eeq
and the result (\ref{30}) follows.

The integrals of the second form (\ref{form2}) $I_3$ and $I_4$
contain an extra factor $q_{1-}$ in the denominator as compared
to $I_1$ and $I_2$. On the formal level this leads to a divergency
of these two integrals at the point $q_{1-}=0$. However in the sum
$I_3 +I_4$ this divergence cancels. Indeed using our approximate
expressions for $D_1$ and $D_2$ valid in the region (\ref{cond1})
we find
\[
I_3+I_4=
\frac{1}{4\pi}\int_{0}^{l_{-}}\frac{dq_{1-}}{q_{1-}}
\Big(-\frac{1}{D_1(-q_{1-})}-\frac{1}{D_2(q_{1-})}\Big)
\]\beq=
\frac{1}{4\pi l_-}\int_{0}^{l_{-}}\frac{dq_{1-}}{q_{1-}}
\Big(
\frac{1}{q_{4+}q_{1-}+(q_4-q_1)_{\perp}^2 +i0}-
\frac{1}{(q_4-q_1)_{\perp}^2-q_{1-}q_{4_+} +i0}\Big)\ .
\label{i34}
\eeq
Obviously the integrand is not singular at $q_{1-}=0$.

We have to integrate this expression in in the small interval around
the points where $D_1=0$ or $D_2=0$. However, as we have seen, the
denominator $D_2$ never vanishes. So in (\ref{i34})
we can drop the second term and in the first term change
\[
\frac{1}{q_{4+}q_{1-}+(q_4-q_1)_{\perp}^2 +i0}\to
-i\pi\delta\Big(q_{4+}q_{1-}+(q_4-q_1)_{\perp}^2)\Big)\ ,
\]
which gives
\beq
I_3+I_4=\frac{i}{4l_- (q_4-q_1)_{\perp}^2}
\label{i34a}
\eeq
that is Eq. (\ref{33a}).

The rest of longitudinal integrals can be  calculated in a similar manner.

Now we are going to demonstrate that one can also calculate our integrals in
the standard manner, factorizing them into two independent ones over $q_{1\pm}$.
Take integral $I_1$. As mentioned one cannot neglect $q_{1+}$ in the first denominator
and $q_{1-}$ in the second without losing convergence. To preserve it we consider
the sum of integrals (\ref{26}) and (\ref{27})
\beq
I_1+I_2=\frac{1}{2i}\int\frac{dq_{1-}}{2\pi}\int\frac{dq_{1+}}{2\pi}
\Big\{\frac{1}{(q_{4}-q_{1})^2+i0}+\frac{1}{(q_4-q_3)^2+i0}\Big\}\frac{1}{(l+q_{1})^{2}+i0}
\ .
\label{i121}
\eeq
Here $q_3=q_4-p_2-q_1$. One observes that convergence in $q_{1-}$ is improved. In order to do this 
with respect to $q_{1+}$ we first pass to integration over $q_3$ with $q_1=q_4-p_2-q_3$
and then rename $q_3\to q_1$ to obtain
\beq
I_1+I_2=\frac{1}{2i}\int\frac{dq_{1-}}{2\pi}\int\frac{dq_{1+}}{2\pi}
\Big\{\frac{1}{(q_{4}-q_{1})^2+i0}+\frac{1}{(q_4-q_3)^2+i0}\Big\}\frac{1}{(l+q_{3})^{2}+i0}
\ .
\label{i122}
\eeq
Taking half the sum of (\ref{i121}) and (\ref{i122}) we finally find
\[
I_1+I_2=\frac{1}{4i}\int\frac{dq_{1-}}{2\pi}\int\frac{dq_{1+}}{2\pi}
\Big\{\frac{1}{(q_{4}-q_{1})^2+i0}+\frac{1}{(q_4-q_3)^2+i0}\Big\}
\]\beq
\times\Big\{\frac{1}{(l+q_{1})^{2}+i0}
+\frac{1}{(l+q_{3})^{2}+i0}\Big\}\ .
\label{i122a}
\eeq 
Now both factors have enough convergence to put $q_{1+}=0$ in the first one and $q_{1-}$ in the second.
The integrals factorizes in two.
\beq
I_1+I_2=\frac{1}{4i}I_+I_-
\label{i123}
\eeq
where
\[
I_-=\int\frac{dq_{1-}}{2\pi}\Big\{\frac{1}{q_{4+}(q_{4-}-q_{1-})+(q_4-q_1)_\perp^2+i0}+
\frac{1}{q_{4+}(p_{1-}+q_{1-})+(q_4-q_3)_\perp^2+i0}\Big\}
\]\beq
=-i\frac{1}{q_{4+}}
\label{i12-}
\eeq
and
\beq
I_+=\int\frac{dq_{1+}}{2\pi}\Big\{\frac{1}{l_-q_{1+}+q_{1\perp}^2+i0}+
\frac{1}{l_-(q_{4+}-p_{2+}-q_{1+})+q_{3\perp}^2+i0}\Big\}=-i\frac{1}{l_-}
\label{i12+}
\eeq
Obviously the result (\ref{i123}) is identical to the the sum of (\ref{i1}) and (\ref{i2})
calculated previously by a different method.

Integrals with $1/q_{1-}$ in the denominator can also be calculated by the standard method
provided one eliminates the singularity at $q_{1-}=0$. In fact the sum of (\ref{28}) and
(\ref{29}) can be rewritten as
\beq
I_3+I_4=\frac{1}{2i}\int\frac{dq_{1-}}{2\pi}\int\frac{dq_{1+}}{2\pi}
\frac{1}{q_{1-}}\frac{1}{(q_{4}-q_{1})^2+i0}\Big\{\frac{1}{(l+q_{1})^{2}+i0}
+\frac{1}{(l+q_{3})^{2}+i0}\Big\}
\ .
\label{i341}
\eeq
where as before $q_3=q_4-p_2-q_1$. Now we can safely put $q_{1+}=0$ in the first factor
and $q_{1-}=0$ in the brackets without losing convergence.
The integral again factorizes in two:
\beq
I_3+I_4=\frac{1}{2i}I_+I_{1-}
\label{i342} 
\eeq
where $I_+$ is the same as before and given by (\ref{i12+}) and
\beq
I_{1-}=\int \frac{dq_{1-}}{2\pi}\frac{1}{q_{1-}}\ \frac{1}{q_{4+}(q_{4-}-q_{1-})+(q_4-q_1)_\perp^2+i0}\ .
\eeq
Here we can safely neglect the term $q_{4+}q_{4-}$ in the denominator since this product is to be
small as compared to squares of the transverse momenta. The singularity at $q_{1-}=0$ then becomes 
spurious. Indeed changing $q_{1-}\to -q_{1-}$ and taking half of the sum we get
\beq
I_{1-}=\frac{1}{2}\int \frac{dq_{1-}}{2\pi}\frac{1}{q_{1-}}\Big\{\frac{1}{-q_{4+}q_{1-}+(q_4-q_1)_\perp^2+i0}-
\frac{1}{q_{4+}q_{1-}+(q_4-q_1)_\perp^2+i0}\Big\}\ .
\label{i34-}
\eeq 
The bracket vanishes at $q_{1-}=0$ so that there is no singularity at this point. Taking the residue in the upper half-plane we find
\beq
I_{1-}=\frac{-i}{2(q_4-q_1)_\perp^2}\ ,
\eeq
so that (\ref{i342})  again coincides with (\ref{i34a}) calculated in our previous manner.

\end{document}